\documentclass[twocolumn,showpacs,pra,superscriptaddress,nofootinbib]{revtex4}
\usepackage{graphicx}
\usepackage{amsmath}

\newcommand{\bra}[1]{\mbox{$\left\langle #1 \right|$}}
\newcommand{\ket}[1]{\mbox{$\left| #1 \right\rangle$}}

\begin{document}

\title{Measurement-device-independent quantum key distribution based on
Bell's inequality}
\author{Hua-Lei Yin}
\author{Yao Fu}
\author{Yan-Lin Tang}
\author{Yuan Li}
\author{Teng-Yun Chen}
\author{Zeng-Bing Chen}
\affiliation{Hefei National Laboratory for Physical Sciences at Microscale and Department
of Modern Physics,\\
University of Science and Technology of China, Hefei, Anhui 230026, China}
\affiliation{CAS Center for Excellence and Synergetic Innovation Center of Quantum Information and Quantum Physics, \\
University of Science and Technology of China, Hefei, Anhui 230026, China \\
}

%%%%%%%%%%%%%%%%%%%%%%%%%%%%%%%%%%%%%%%%%%%%%%%%%%%%%%%%%%%%%%%%%%%%%%%
% Abstract
%%%%%%%%%%%%%%%%%%%%%%%%%%%%%%%%%%%%%%%%%%%%%%%%%%%%%%%%%%%%%%%%%%%%%%%

\begin{abstract}
We propose two quantum key distribution (QKD) protocols based on Bell's
inequality, which can be considered as modified time-reversed E91 protocol.
Similar to the measurement-device-independent quantum key distribution
(MDI-QKD) protocol, the first scheme requires the assumption that Alice and
Bob perfectly characterize the encoded quantum states. However, our second
protocol does not require this assumption, which can defeat more known and
unknown source-side attacks compared with the MDI-QKD. The two protocols are
naturally immune to all hacking attacks with respect to detections.
Therefore, the security of the two protocols can be proven based on the
violation of Bell's inequality with measurement data under fair-sampling
assumption. In our simulation, the results of both protocols show that
long-distance quantum key distribution over 200 km remains secure with
conventional lasers in the asymptotic-data case. We present a new technique
to estimate the Bell's inequality violation, which can also be applied to
other fields of quantum information processing.
\end{abstract}

\maketitle

\section{INTRODUCTION}

Quantum key distribution (QKD), such as BB84 \cite{BB_84} and E91 \cite%
{ekert1991quantum}, provides a secure way to exchange private information.
It enables a common string of random bits, called secret keys, to be shared
secretly between the two legitimate users (typically called Alice and Bob).
In principle, QKD exploits the fundamental laws of quantum mechanics to
offer information-theoretical security \cite%
{RevModPhys:02:Gisin,RevModPhys:09:Scarani}. However, the gap between the
ideal devices fulfilling the assumptions of security proof and the realistic
ones opens various loopholes which make the system suffered from various
kinds of side-channel attacks \cite%
{zhao:Quantum:2008,xu:2010:experimental,Lydersen:BrightAttack:2010,weier:2011:quantum,gerhardt:2011:full}%
.

In general, there are two approaches to circumvent the side-channel attacks.
The first one is trying to characterize realistic devices completely in the
security proofs. %and account for all side channels.
This approach is quite difficult since it is almost impossible to have a
special model that includes all practically relevant imperfections of
realistic devices. The second one is known as (full) device-independent QKD
(DI-QKD) \cite{pironio:2009:device,masanes:2011:secure} whose security proof
is based on the observation of nonlocal statistical correlations
(loophole-free test of Bell's inequality) only and as such, it does not
require detailed knowledge of the devices.
%Inspired by the time-reversed BB84 protocol and combing with a self-testing scheme,
A recent DI-QKD protocol has been proposed \cite{Lim:2013:Device}, where the
violation of loophole-free Bell's inequality is not affected by the channel
losses between Alice and Bob, because it only requires Bell test performed
locally in Alice's site. Unfortunately, DI-QKD is currently highly
impractical, for the reason that it requires the legitimate users to carry
out a (full) loophole-free Bell test (very high detection efficiency and
space-like separation between Alice and Bob), which is still a big
experimental challenge even with the state-of-the-art technologies \cite%
{giustina:2013:bell,Christensen:2013:Detection}. More importantly, its
secure key rate is very limited at practical distances even using the novel
techniques, i.e., local Bell test \cite{Lim:2013:Device} or heralded qubit
amplifier \cite{Gisin:2010:Proposal}.

Recent progress has been made by introducing the novel idea of
measurement-device-independent QKD (MDI-QKD) protocol \cite{Lo:MIQKD:2012},
which is built on the idea of the time-reversed Einstein-Podolsky-Rosen
protocol for QKD \cite{Biham:1996:Quantum,Inamori:TimeReverseEPR:2002}. The
measurement devices in MDI-QKD, which can be treated as a true black box,
are essentially used to post-select entanglement states from the mixed
states between Alice and Bob. Thus, MDI-QKD closes all kinds of
detection-side loopholes. Furthermore, one crucial advantage of the MDI-QKD
is that the encoded quantum states can use weak coherent pulses (WCPs)
combined with the decoy-state techniques \cite%
{Hwang:2003:Quantum,Lo:2005:Decoy,Wang:2005:Beating} instead of
single-photon sources. Besides, the secure key rate and transmission distance are
comparable to that of usual QKD protocols with entangled sources \cite%
{Ursin:2007:Eantanglement,Ma:2007:quantum}. An important assumption in
MDI-QKD is that Alice and Bob need to perfectly characterize the encoded
quantum states.
The secret key distribution of BB84 protocol is based on information encoded
complementary bases, while the secret key distribution of E91 is based on
quantum entanglement.
The E91 protocol is the first QKD scheme whose security proof exploits
the violation of Bell's inequality. As a security assumption of usual QKD,
Alice and Bob need to trust their devices (both source-side and
detection-side), the Bell test can then be performed with the
measurement data under the fair-sampling assumption.

In this paper, we propose two QKD protocols based on Bell's inequality,
which can be regarded as the modified time-reversed E91 protocol, denoted by
P1 and P2.
The two protocols are naturally immune to all possible detection-side
attacks. P1 requires the assumption that Alice and Bob need to perfectly
characterize the encoded quantum states. However, P2 does not require this
assumption. Therefore, P2 is more device-independent, which enables the
system to defeat more known and unknown source-side attacks compared with
the MDI-QKD. In contrast to DI-QKD, the two schemes proposed here do not
require the legitimate users to perform a loophole-free Bell test. It is
enough to prove our two schemes' security based on the violation of Bell's
inequality with measurement data under fair-sampling assumption. We
demonstrate that P1 is equivalent to the MDI-QKD protocol in the asymptotic case. Combining the conventional laser
sources with vacuum+decoy+signal method, we simulate the secure key rates in
the asymptotic-data case and the finite-data case, respectively. The results
of both protocols show that long-distance quantum key distribution over 200
km remains secure with conventional lasers in the asymptotic-data case. We
present a new technique to estimate the violation of Bell's inequality (the
\textquotedblleft Bell value\textquotedblright ), which can be used to test
local realism without preparing entanglement states in advance.

\section{NECESSARY ASSUMPTIONS}

\label{Section:necessary:assumptions} For each QKD protocol, the security
assumptions play a crucial role. In order to show our QKD protocols
sufficiently, we first illustrate five fundamental assumptions of P1 and P2,
which are also necessary in DI-QKD protocol \cite%
{masanes:2011:secure,Lim:2013:Device}.

First, Alice and Bob's physical locations are isolated and secure, i.e., no
unwanted information can leak out from the secure location. Second, they
trust their quantum random number generators to generate a random
output. Third, they can compute and store the classical data with their
trusted classical devices. Fourth, the two legitimate users could share an
authenticated classical channel. Fifth, the (quantum) devices of different
users are causally independent.
The last assumption is guaranteed when the devices'
memory is totally erased after each process or the devices have no internal
memory at all (this assumption is necessary for defeating memory attack \cite%
{Barrett:2013:Memory}).

In addition to the above assumptions, the security of P1 and MDI-QKD will be
guaranteed with another two assumptions. The first one is that the Hilbert
space of quantum state preparation is two-dimensional. The second assumption
is that Alice and Bob can perfectly characterize their encoded quantum
states (e.g., the polarization encoded scheme of phase-randomized WCPs).
Nevertheless, without the second security assumption, P2 still satisfy the
security proof. Thus, P2 can defeat more known and unknown source-side
attacks.

Note that it is also not required Alice and Bob to characterize their encoded quantum
states perfectly in recent works \cite{Brunner:2011:Semi,Yin:2013:Measurement,Li:2014:Quantum}, but the single-photon
sources assumption is necessary.
In our scheme, we use conventional laser sources (WCPs) which make our QKD protocols more practical and
economical under current technology instead of single-photon sources.

\section{PROTOCOL DESCRIPTION}

In the following, we describe the QKD schemes in details, see Fig.~\ref%
{Fig:Setup}.
\begin{figure}[tbh]
\centering
\resizebox{8.5cm}{!}{\includegraphics{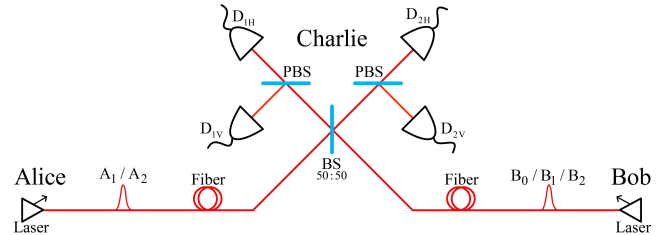}}
\caption{(Color online) Basic setup of P1 and P2 protocols. For simplicity,
we consider the polarization encoding scheme. Alice (Bob) randomly prepares
two (three) settings \{$A_{1},A_{2}$\} (\{$B_{0},B_{1},B_{2}$\}) of quantum
states with phase randomized WCPs. Charlie performs Bell state measurement
and the measurement results are publicly announced. A successful Bell state
measurement corresponds to the observation of only two of four detectors
being clicked. $\mbox{$\left| \psi^+ \right\rangle$}=1/\protect\sqrt{2}(
\mbox{$\left| HV \right\rangle$}+\mbox{$\left| VH \right\rangle$}$
represents a click in $D_{1H}$ and $D_{1V}$, or $D_{2H}$ and $D_{2V}$, while
$\mbox{$\left| \psi^- \right\rangle$}=1/\protect\sqrt{2}(\mbox{$\left| HV
\right\rangle$}-\mbox{$\left| VH \right\rangle$})$ represents a click in $%
D_{1H}$ and $D_{2V}$, or $D_{2H}$ and $D_{1V}$. }
\label{Fig:Setup}
\end{figure}
Alice and Bob independently and randomly prepare quantum states with phase
randomized WCPs in two settings \{$A_{1}=\sigma _{z},A_{2}=\sigma _{x}$\}
and three settings \{$B_{0}=\sigma _{z},B_{1}=(\sigma _{z}+\sigma _{x})/%
\sqrt{2},B_{2}=(\sigma _{z}-\sigma _{x})/\sqrt{2}$\}, respectively.
Then they send each pulse to an untrusted third party Charlie, who can be
anybody, even the eavesdropper Eve. Charlie carries out a partial Bell state
measurement (BSM). As is known, we cannot completely distinguish four Bell
states simultaneously through singly using linear optical element. In this
paper we can only unambiguously distinguish two Bell states \{$\mbox{$\left|
\psi^+ \right\rangle$}$,$\mbox{$\left| \psi^- \right\rangle$}$\}
(Fortunately, the identification of one Bell state is adequate to prove
security). Charlie announces through a public channel whether he has
received a Bell state and which Bell state he has received. Alice and Bob
keep the raw data of successful BSM results and discard the rest. The Bell
value can be estimated from the raw data of quantum states sent by Alice's
and Bob's two settings (bases) \{$A_{1},A_{2}$\} and \{$B_{1},B_{2}$\},
respectively. They post-select the results as a raw key when Alice and Bob
choose setting $A_{1}$ and $B_{0}$, respectively (here, $A_{1}=B_{0}=\sigma
_{z}$). Decoy-state techniques are employed \cite%
{Hwang:2003:Quantum,Lo:2005:Decoy,Wang:2005:Beating} to estimate the yield,
bit error rate and Bell value, given that both Alice and Bob
send out single-photon states (untagged portion). One party needs to carry
out a bit flip to his or her raw data to guarantee that their raw key is
correctly correlated. Then they perform error-correction and privacy
amplification with one-way classical postprocessing to extract secure keys.

\section{SECURITY ANALYSIS}

In this section, we present a brief description of P1's and P2's security
against collective attacks and the main results of secure key rate. Here, we
focus on collective attacks where Eve adopts the same attack to each system
of Alice and Bob. For the first QKD protocol, P1, only signals originated
from single-photon pulses emitted by both Alice and Bob are guaranteed to be
secure while Eve's information is restricted by the Holevo bound \cite%
{RevModPhys:09:Scarani,pironio:2009:device}.
Since the WCPs' phase randomization makes the emitted quantum states of
Alice and Bob into a classical mixture of states, it enables Alice and Bob
to tag each pulse in principle though they do not need to do so in practice
\cite{GLLP:2004:Security}. It is assumed that Eve competely knows the
information from the multiphoton components (tagged portion). Then the
information of Eve is composed of two portions, namely tagged and untagged portion,
which can be written as (see Appendix~\ref{Appsection:HOLEVOBOUND} for more
details)
\begin{equation}
\begin{aligned} \chi_{1}(A_{1}:E)&=\chi_{1}^{\rm tag}(A_{1}:E)+\chi_{1}^{\rm
untag}(A_{1}:E)\\
&=(Q_{\mu\nu}^Z-Q_{11}^Z)+Q_{11}^ZH\left(e_{11}^{BZ}+\frac{S_{11}}{2%
\sqrt{2}}\right). \end{aligned}  \label{I:A:E}
\end{equation}%
The mutual information between Alice and Bob, considering that the
error-correction will leak extra information, is given by
\begin{equation}
\begin{aligned}
I_{1}(A_{1}:B_{0})=Q_{\mu\nu}^Z-Q_{\mu\nu}^ZfH(E_{\mu\nu}^Z). \end{aligned}
\label{I:A:B}
\end{equation}%
The secure key rate of P1 (per joint signal state emitted by Alice and Bob
simultaneously in $\sigma _{z}$ basis) can be written as
\begin{equation}
\begin{aligned} R_{1} &=I_{1}(A_{1}:B_{0})-\chi_{1}(A_{1}:E)\\ & =
Q_{11}^Z\Big[1-H\left(e_{11}^{BZ}+\frac{S_{11}}{2\sqrt{2}}\right)\Big]-Q_{%
\mu\nu}^ZfH(E_{\mu\nu}^Z), \end{aligned}  \label{P1}
\end{equation}%
where $Q_{\mu \nu }^{Z}$ and $E_{\mu \nu }^{Z}$, the overall gain and
quantum bit error rate (QBER), can be directly obtained from the
experimental results. The subscript $\mu \nu $ means that Alice and Bob send
out WCPs with intensity $\mu $ and $\nu $, respectively. For
the single-photon states, the gain $Q_{11}^{Z}$, bit error rate $e_{11}^{BZ}$
and the Bell value $S_{11}$ can be estimated by the decoy-state method.
Here, the parameter $f$ is the error correction efficiency (we take the value $%
f=1.16$ in our simulation), and $H(e)=-e\log _{2}(e)-(1-e)\log _{2}(1-e)$ is
the binary Shannon entropy function.

For QKD protocol P2, the multiphoton components are tagged whose
information will be fully leaked to Eve \cite{GLLP:2004:Security}. Only
signals originated from single-photon pulses emitted by both Alice and Bob
are the untagged portion which can be extracted as secure keys. For the
untagged portion, we use the min-entropy to bound Eve's knowledge of the
secure keys, which has been applied to analyze security in Refs.~\cite%
{masanes:2011:secure,Li:2014:Quantum}. Details of this part can be found in
Appendix~\ref{Appsection:Min-entropy}. The secure key rate of P2 is given by
\begin{equation}
\begin{aligned} R_{2}&=I_{2}(A_{1}:B_{0})-\chi_{2}(A_{1}:E)\\ & =
Q_{11}^{Z}\Big[1-{\rm
log_{2}}\left(1+\sqrt{2-\frac{S_{11}^2}{4}}\right)\Big]-Q_{\mu\nu}^ZfH(E_{%
\mu\nu}^Z). \end{aligned}  \label{P2}
\end{equation}%
The second term $Q_{\mu \nu }^{Z}fH(E_{\mu \nu }^{Z})$ quantifies the amount
of information needed for the error-correction. The non-trivial part of our
bound is $\mathrm{log_{2}}\left( 1+\sqrt{2-S_{11}^{2}/{4}}\right) Q_{11}^{Z}$%
, which quantifies Eve's information.

When the phases of the WCPs sent by Alice and Bob are fully randomized,
the density matrix of the quantum states should be written as
\begin{equation}
\begin{aligned} \rho = \int_0^{2\pi} \frac{d\theta}{2\pi}
|\sqrt{\mu}e^{i\theta}\rangle\langle
\sqrt{\mu}e^{i\theta}|=e^{-\mu}\sum_{n=0}^\infty \frac{\mu^n}{n!}
|n\rangle\langle n|,\\ \end{aligned}
\label{density mateix of coherent state}
\end{equation}%
where $\theta $ and $\mu $ are the phase and intensity of the coherent
states, respectively. Then the quantum channel can be considered as a photon
number channel \cite{Lo:2005:Decoy}. The overall gain and QBER in $\sigma
_{z}$ basis can be given by
\begin{equation}
\begin{aligned} Q_{\mu\nu}^{Z} &=
Q_{\mu\nu}^{CZ}+Q_{\mu\nu}^{EZ}=\sum_{n=0}^\infty\sum_{m=0}^\infty\frac{%
\mu^n\nu^m}{n!m!}e^{-\mu-\nu}Y_{nm}^{Z},\\ E_{\mu\nu}^{Z}Q_{\mu\nu}^{Z}&=
e_{d}Q_{\mu\nu}^{CZ}+(1-e_{d})Q_{\mu\nu}^{EZ}\\
&=\sum_{n=0}^\infty\sum_{m=0}^\infty\frac{\mu^n\nu^m}{n!m!}e^{-\mu-%
\nu}e_{nm}^{BZ}Y_{nm}^{Z}, \end{aligned}  \label{QZ}
\end{equation}%
where $Y_{nm}^{Z}$ ($e_{nm}^{BZ}$) is the yield (bit error rate), given that
Alice and Bob send out $n$-photon and $m$-photon pulse, respectively. $%
Q_{\mu \nu }^{CZ}$ ($Q_{\mu \nu }^{EZ}$) is the total gain of a successful
BSM when the polarization of the pulses sent by Alice and Bob are different
(the same) in $\sigma _{z}$ basis, which represents a correct (false)
measurement result. $e_{d}$ represents the overall misalignment-error probability of the system.
The Bell value $S_{11}$ is given by
\begin{equation}
\begin{aligned}
S_{11}&=\frac{1}{2}(S_{11}^{\psi^-}+S_{11}^{\psi^+})=S_{11}^{\psi^-},\\
S_{11}^{\psi^-}&=\langle A_{2}B_{2}\rangle_{11}^{\psi^-}-\langle
A_{2}B_{1}\rangle_{11}^{\psi^-}-\langle
A_{1}B_{2}\rangle_{11}^{\psi^-}-\langle A_{1}B_{1}\rangle_{11}^{\psi^-},
\end{aligned}  \label{CHSH}
\end{equation}%
where we use $S_{11}^{\psi ^{-}}=S_{11}^{\psi ^{+}}$ because of symmetry. In
our simulation, the expectation of single-photon states $\langle
A_{k}\otimes B_{l}\rangle _{11}^{\psi ^{-}}=\langle A_{k}B_{l}\rangle
_{11}^{\psi ^{-}}$ results from the successful projection into the Bell
state $\mbox{$\left| \psi^- \right\rangle$}$ with appropriate setting of $%
A_{k}$ and $B_{l}$, where $k,l\in \{1,2\}$. So the expectation is
given by
\begin{equation}
\begin{aligned} \langle A_{k}B_{l}\rangle_{11}^{\psi^-}=&(1-2e_{d})\\
&\times\frac{Y_{H_{A_{k}}H_{B_{l}}}^{11\psi^-}+Y_{V_{A_{k}}V_{B_{l}}}^{11%
\psi^-}-Y_{H_{A_{k}}V_{B_{l}}}^{11\psi^-}-Y_{V_{A_{k}}H_{B_{l}}}^{11%
\psi^-}}{Y_{H_{A_{k}}H_{B_{l}}}^{11\psi^-}+Y_{V_{A_{k}}V_{B_{l}}}^{11%
\psi^-}+Y_{H_{A_{k}}V_{B_{l}}}^{11\psi^-}+Y_{V_{A_{k}}H_{B_{l}}}^{11%
\psi^-}}, \end{aligned}  \label{EAB}
\end{equation}%
where $Y_{H_{A_{k}}V_{B_{l}}}^{11\psi ^{-}}$ is a yield. The superscript $%
11\psi ^{-}$ represents that Charlie obtains a Bell state $\ket{\psi^{-}}$
successfully, given that both Alice and Bob send out single-photon states.
The subscript $H_{A_{k}}V_{B_{l}}$ represents the joint quantum state that
Alice sends out a positive eigenvalue corresponding to the eigenstate of
setting $A_{k}$ while Bob sends out a negative eigenvalue corresponding to
the eigenstate of setting $B_{l}$.

We present two methods to obtain $Y_{11}^{Z}$, $e_{11}^{BZ}$ and $S_{11}$,
the relevant parameters which are needed to evaluate the key rate formula
above, given that Alice and Bob send Charlie a finite number of signals and
use a finite number of decoy states. We use the
standard error analysis method \cite{Ma:2005:Practical,Ma:2012:Statistical} to solve this problem
(a rigorous estimation can be acquired by using large deviation theory, i.e.,
the Chernoff bound \cite{curty:2014:finite}). More precisely, we combine
linear programming and analytical method, respectively, with two decoy
states, to estimate all the lower bounds of $Y_{11}^{Z}$, $e_{11}^{BZ}$ and $%
S_{11}$ within single-photon states. Importantly, our methods are valid for
arbitrary photon-number distribution of signals sent by Alice and Bob. To
get more details of this part, please see Appendix~\ref{Appsection:Estimate}.

\section{SIMULATION RESULTS}

In this section, we analyze the behavior of the secret key rates of P1 and
P2 provided in Eq.~\eqref{P1} and Eq.~\eqref{P2}, respectively. In our
simulation, the loss of fiber-based channel is 0.2 dB/km. For simplicity, we
assume that all detectors are identical (i.e., they have the same detection
efficiency and background count rate), and their background count rate, to a
good approximation, is independent of incoming signals. We assume that the
detection efficiency of Charlie is 40\% and the background count rate is $%
3\times 10^{-6}$. We use an intrinsic error rate that represents the
misalignment and instability of the optical system. Furthermore, the
security bound is fixed to be $\epsilon =10^{-10}$.

\begin{figure}[tbh]
\centering
\resizebox{8.5cm}{!}{\includegraphics{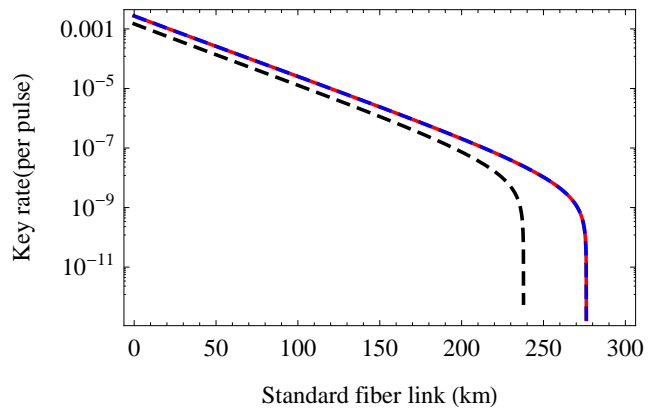}}
\caption{(Color online) The secure key rates in asymptotic case. Asymptotic
case means that Alice and Bob use infinite number of decoy states and send
Charlie infinite data signals. We use the following practical experimental
parameters: the detection efficiency $\protect\eta_{d}$ of Charlie is 40\%,
the intrinsic loss coefficient $\protect\beta$ of the standard telecom fiber
channel is 0.2 dB/km, the overall misalignment-error probability $e_{d}$ of
the system is 1.5\%, the background count rate $p_{d}$ is $3\times10^{-6}$,
the intensity of signal state $\protect\mu$ is 0.3.}
\label{Fig:keyrate1}
\end{figure}

The secure key rates of P1 and P2 in the asymptotic case are shown in Fig.~%
\ref{Fig:keyrate1} with blue dashed curve and black dashed curve,
respectively. Meanwhile, we also present the simulation result of the
MDI-QKD \cite{Lo:MIQKD:2012} with the red solid curve. We can see clearly
that the secure key rate and secure distance of P1 are the same as MDI-QKD's
in the asymptotic case. The reason lies in that the security proof based on
entanglement distillation purification is equivalent to direct
information-theoretic arguments with one-way classical communications. The
secure key rate and secure distance of P2 are both less than P1's, since P2
requires fewer security assumptions, i.e., we do not require that Alice and
Bob perfectly characterize their encoded quantum states.

\begin{figure}[tbh]
\centering
\resizebox{8.5cm}{!}{\includegraphics{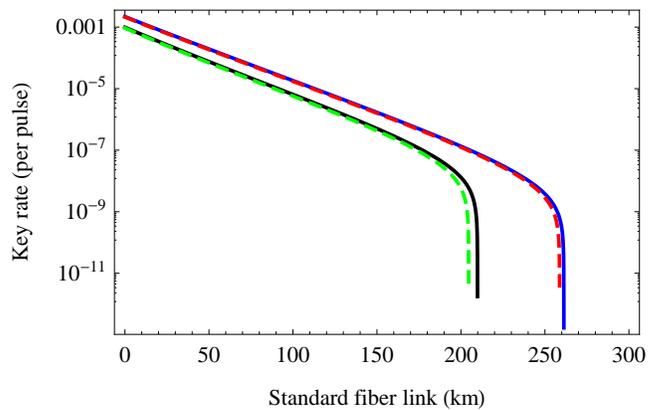}}
\caption{(Color online) The secure key rates with two decoy states in
asymptotic-data case. The intensities of signal state $\protect\mu $ and one
decoy state $\protect\nu $ are 0.3 and 0.01, respectively, while the other
decoy state is a vacuum state. We emphasize that the key rates with
analytical method of Appendix~\ref{Appsection:Estimate} almost
overlap with the one with linear programming, which shows that the
analytical method provides an excellent estimation. The estimation using two
decoy states gives a secure key rate which is nearly the same as the one
using infinite decoy states. Therefore, two decoy states (vacuum+decoy) are
enough for a near-optimal estimation, no matter how many decoy states are
added, the secure key rate cannot be improved too much. In the
asymptotic-data and two decoy states case, the security distances of P1 and
P2 are more than 200 km.}
\label{Fig:keyrate2}
\end{figure}
\begin{figure}[tbh]
\centering
\resizebox{8.5cm}{!}{\includegraphics{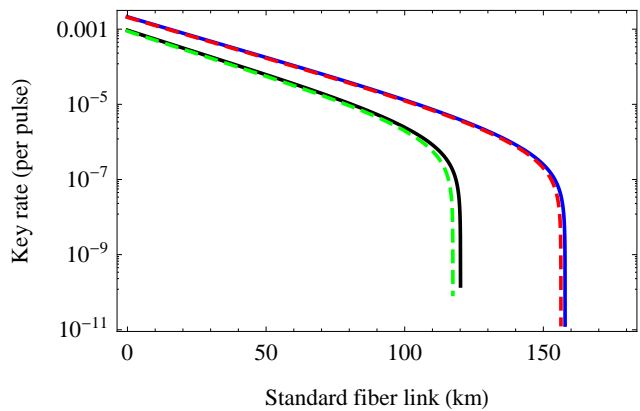}}
\caption{(Color online) The secure key rates with statistical fluctuations.
The intensities of signal state $\protect\mu $ and one decoy state $\protect%
\nu $ are 0.3 and 0.01, respectively, while the other decoy state is a
vacuum state. The finite data is $N=10^{14}$, the secure bound is $\protect%
\epsilon =10^{-10}$. In the finite-data and two decoy states case, the
security distance of P1 is more than 150 km, and the security distance of P2
is more than 110 km.}
\label{Fig:keyrate3}
\end{figure}

In practice, we need to consider a finite number of decoy states.
The simulation results using linear programming and analytical method with
vacuum+decoy states in asymptotic-data case (finite-data case) are shown in
Fig.~\ref{Fig:keyrate2} (Fig.~\ref{Fig:keyrate3}). Notice that the key rates
using the analytical method almost overlap with the one using linear
programming in Fig.~\ref{Fig:keyrate2} and Fig.~\ref{Fig:keyrate3}. In the
asymptotic-data case (Fig.~\ref{Fig:keyrate2}), the blue (black) solid curve
represents the secure key rate of P1 (P2) under linear programming, while
the red (green) dashed curve represents the secure key rate of P1 (P2) under
analytical method. Comparing Fig.~\ref{Fig:keyrate1} with Fig.~\ref%
{Fig:keyrate2}, we can see clearly that the key rates with two decoy states
(vacuum+decoy) are close to the corresponding ones with infinite number of decoy states.
In finite-data case (Fig.~\ref{Fig:keyrate3}), the statistical fluctuations
are simulated using the standard error analysis method \cite%
{Ma:2005:Practical}. For simplicity, we assume that Alice and Bob send same
number of pulses for all $\mu _{A_{k}}\uplus \nu _{B_{l}}$ channels, denoted
by $N$ (an efficient parameter optimization method can be found in \cite%
{Xu:2014:Protocol}). Here $\mu _{A_{k}}\uplus \nu _{B_{l}}$ is defined as
the case that Alice sends out WCPs of intensity $\mu $ with setting $A_{k}$
while Bob sends out WCPs of intensity $\nu $ with setting $B_{l}$, where $%
k\in \{0,1\}$, $l\in \{0,1,2\}$. In the finite-data and two decoy states
cases, the security distance of P1 (P2) is more than 150 km (110 km).

\section{CONCLUSION}

In summary, we have proposed two QKD protocols, P1 and P2, inspired by E91
and MDI-QKD protocols. As to P1, the security assumptions and the secure key
rate in asymptotic case are the same as MDI-QKD's. More importantly, in the
security proof of P2, Alice and Bob's perfectly characterizing encoded
quantum states is not required. Thus, P2 is more resistant to source-side
attacks compared with MDI-QKD. The simulation results show that P2 is more
practical using conventional laser sources and decoy-state method instead of the single-photon
sources. P2 depends less on device but keeps a high secure key rate and long
transmission distance. Moreover, the Bell value can be estimated accurately
with conventional laser sources and finite-number decoy states method.
We believe that this technique can be used in other fields of quantum
information processing. The full parameter optimization of P1 and P2 needs
to be done in the future.

\section{ACKNOWLEDGMENTS}

This work was supported by the NNSF of China under Grant No. 61125502, the
National Fundamental Research Program under Grant No. 2011CB921300, the CAS
and the National High Technology Research and Development Program of China.

\begin{appendix}

\section{HOLEVO BOUND} \label{Appsection:HOLEVOBOUND}
Without loss of generality, the BB84 protocol implies that one can compute the bound by restricting consideration to collective attacks \cite{RevModPhys:09:Scarani}. Considering the collective attacks, the final density matrix of Alice and Bob's joint quantum state can be given by
\begin{equation} \label{density:matrix:AB}
\begin{aligned}
\rho_{AB}=&\lambda_{1}\ket{\phi^+}\bra{\phi^+}+\lambda_{2}\ket{\phi^-}\bra{\phi^-}\\
&+\lambda_{3}\ket{\psi^+}\bra{\psi^+}+\lambda_{4}\ket{\psi^-}\bra{\psi^-},
\end{aligned}
\end{equation}
with $\sum_{i=1}^4 \lambda_{i}=1$. The four Bell states
\begin{equation} \label{Bell:state}
\begin{aligned}
\ket{\phi^+}=\frac{1}{\sqrt{2}}(\ket{HH}+\ket{VV})=\frac{1}{\sqrt{2}}(\ket{++}+\ket{--}),\\
\ket{\phi^-}=\frac{1}{\sqrt{2}}(\ket{HH}-\ket{VV})=\frac{1}{\sqrt{2}}(\ket{+-}+\ket{-+}),\\
\ket{\psi^+}=\frac{1}{\sqrt{2}}(\ket{HV}+\ket{VH})=\frac{1}{\sqrt{2}}(\ket{++}-\ket{--}),\\
\ket{\psi^-}=\frac{1}{\sqrt{2}}(\ket{HV}-\ket{VH})=\frac{1}{\sqrt{2}}(\ket{-+}-\ket{+-}),
\end{aligned}
\end{equation}
constitute a complete orthogonal basis in two-dimensional Hilbert space.
$\ket{\phi^{\pm}}$ ($\ket{\phi^+},\ket{\psi^+}$) are perfectly correlated in $\sigma_z$ ($\sigma_x$)
basis, while $\ket{\psi^{\pm}}$ ($\ket{\phi^-},\ket{\psi^-}$) are perfectly anticorrelated. Therefore, the bit error rates in $\sigma_z$ and $\sigma_x$ basis are given by
\begin{equation} \label{Bit:error}
\begin{aligned}
e^{BZ}=\lambda_{3}+\lambda_{4}, \quad e^{BX}=\lambda_{2}+\lambda_{4}.
\end{aligned}
\end{equation}
The phase error rates in the two bases are
\begin{equation} \label{Bit:error}
\begin{aligned}
e^{PZ}=\lambda_{2}+\lambda_{4}=e^{BX}, \quad e^{PX}=\lambda_{3}+\lambda_{4}=e^{BZ}.
\end{aligned}
\end{equation}
The secure key rate of the entanglement distillation purification-based QKD using one-way classical communications is \cite{BDSW:1996:Mixed,Shor:2000:Simple}
\begin{equation} \label{EDP:keyrate}
\begin{aligned}
R_{EDP}=1-H(e^{BZ})-H(e^{PZ})=1-H(e^{BZ})-H(e^{BX}).
\end{aligned}
\end{equation}

Here, we use Holevo bound to estimate Eve's information \cite{Kraus:2005:Lower,Renner:2005:Information},
\begin{equation} \label{Holevo:bound:Eve}
\begin{aligned}
\chi(A:E)&=S(\rho_{E})-\frac{1}{2}S(\rho_{E|0})-\frac{1}{2}S(\rho_{E|1})\\
&=H(e^{BX}),
\end{aligned}
\end{equation}
and the secure key rate is
\begin{equation} \label{Holevo:bound:keyrate}
\begin{aligned}
R_{Inf}&=I(A:B)-\chi(A:E)\\
&=1-H(e^{BZ})-H(e^{BX})=R_{EDP}.
\end{aligned}
\end{equation}
We can see that the security proof based on entanglement distillation purification is equivalent to direct information-theoretic arguments with one-way classical communications.

The Bell operator can be written as
\begin{equation} \label{CHSH}
\begin{aligned}
B&=A_{1}\otimes B_{1} + A_{1}\otimes B_{2} + A_{2}\otimes B_{1} -A_{2}\otimes B_{2}\\
&=\sqrt{2}(\sigma_{z}\otimes \sigma_{z}+\sigma_{x} \otimes \sigma_{x}).
\end{aligned}
\end{equation}
Thus, the Bell value is given by \cite{pironio:2009:device}
\begin{equation} \label{CHSH:value}
\begin{aligned}
S&=Tr(B\rho_{AB})\\
&=\sqrt{2}Tr((\sigma_{z}\otimes \sigma_{z}+\sigma_{x} \otimes \sigma_{x})\rho_{AB})\\
&=2\sqrt{2}(\lambda_{1}-\lambda_{4})\\
&=2\sqrt{2}(1-e^{BZ}-e^{BX}).
\end{aligned}
\end{equation}
Instead of using the bit error rate $e^{BX}$ in $\sigma_x$ basis, the parameter from which Eve's information is inferred is the average Bell value $S$ and the bit error rate $e^{BZ}$ in $\sigma_z$ basis, i.e., $e^{BX}=1-e^{BZ}-S/2\sqrt{2}$.

Consider that Alice and Bob encode their bits in the polarization degrees of freedom of phase-randomized WCPs. The information of Eve with two portions \cite{GLLP:2004:Security}, i.e., tagged and untagged  portion, can be written as
\begin{equation} \label{I:A:E}
\begin{aligned}
\chi_{1}(A_{1}:E)&=\chi_{1}^{\rm tag}(A_{1}:E)+\chi_{1}^{\rm untag}(A_{1}:E)\\
&=(Q_{\mu\nu}^Z-Q_{11}^Z)+Q_{11}^ZH\left(1-e_{11}^{BZ}-\frac{S_{11}}{2\sqrt{2}}\right)\\
&=(Q_{\mu\nu}^Z-Q_{11}^Z)+Q_{11}^ZH\left(e_{11}^{BZ}+\frac{S_{11}}{2\sqrt{2}}\right),
\end{aligned}
\end{equation}
where the superscripts $tag$ and $untag$ represent tagged portion and untagged portion, respectively. The mutual information between Alice and Bob, considering that the error-correction will leak extra information, is given by
\begin{equation} \label{I:A:B}
\begin{aligned}
I_{1}(A_{1}:B_{0})=Q_{\mu\nu}^Z-Q_{\mu\nu}^ZfH(E_{\mu\nu}^Z).
\end{aligned}
\end{equation}
Finally, the secure key rate of P1 is given by
\begin{equation} \label{finally key rate 1}
\begin{aligned}
R_{1} &=I_{1}(A_{1}:B_{0})-\chi_{1}(A_{1}:E)\\
& = Q_{11}^Z\Big[1-H\left(e_{11}^{BZ}+\frac{S_{11}}{2\sqrt{2}}\right)\Big]-Q_{\mu\nu}^ZfH(E_{\mu\nu}^Z).
\end{aligned}
\end{equation}

\section{MIN-ENTROPY} \label{Appsection:Min-entropy}
In this part, the goal is to guarantee the security proof of P2 although removing the assumption that encoded quantum states need to be characterized perfectly. Obviously, the first five assumptions in section \ref{Section:necessary:assumptions} are also required in the security proof of DI-QKD.
The secure key rate of DI-QKD \cite{masanes:2011:secure} is
\begin{equation} \label{DI-QKD:keyrate}
\begin{aligned}
R=H_{\rm min}^{\rm DI}(A_{1}|E)-H_{\rm con}^{\rm DI}(A_{1}|B_{0}),
\end{aligned}
\end{equation}
where
\begin{equation} \label{DI-QKD:keyrate1}
\begin{aligned}
H_{\rm min}^{\rm DI}(A_{1}|E)&=-{\rm log_{2}}P_{\rm guess}(a),\\
H_{\rm con}^{\rm DI}(A_{1}|B_{0})&=H(e^{BZ}).
\end{aligned}
\end{equation}
In above equations, $H_{\rm min}^{\rm DI}(A_{1}|E)$ is the (quantum) min-entropy, which will be used for restricting the knowledge of Eve. By employing privacy amplification, we are able to make Eve's information arbitrarily small. $H_{\rm con}^{\rm DI}(A_{1}|B_{0})$ is the conditional Shannon entropy which quantifies the amount of information needed for error-correction. $a$ is the output (eigenvalue) of setting $\{A_{1},A_{2}\}$, and $P_{\rm guess}(a)$ is the maximal guessing probability which is used for quantifying the degree of unpredictability of Alice's measurement output $a$. The following bound will hold in Bell's inequality \cite{pironio:2010:random}
\begin{equation} \label{guessing:probability}
\begin{aligned}
P_{\rm guess}(a)\leq\frac{1}{2}+\frac{1}{2}\sqrt{2-\frac{S^2}{4}}.
\end{aligned}
\end{equation}
In the DI-QKD scheme, the loophole-free Bell test can ensure QKD security against untrusted detectors and arbitrarily dimensional quantum systems. P2 can be regarded as the modified time-reversed E91 and it is naturally immune to all possible detection-side attacks. The quantum states of P2 are required to be prepared in the two-dimensional Hilbert space, because the security of high-dimensional quantum states will not be guaranteed  (for example, the four-dimensional separable state will have the property of two-dimensional maximally entangled state in Ref.~\cite{pironio:2009:device}). Therefore, we can use the measurement data to calculate the Bell value with the assumption that the Hilbert space of quantum state preparation is two-dimensional.  We use the min-entropy to bound Eve's information with the untagged portion
\begin{equation} \label{min-entropy:untagged}
\begin{aligned}
\chi_{2}^{\rm untag}(A_{1}:E)&=Q_{11}^{Z}\big[1-H_{\rm min}^{\rm 2\dim}(A_{1}|E)\big]\\
&\leq Q_{11}^{Z}\big[1+{\rm log_{2}}\left(\frac{1}{2}+\frac{1}{2}\sqrt{2-\frac{S_{11}^2}{4}}\right)\big],
\end{aligned}
\end{equation}
where the superscript $\rm 2\dim$ represents that the Hilbert space of quantum systems is two-dimensional. From the analysis above, it is not necessarily required that Alice and Bob perfectly characterize their encoded quantum states. Eve will acquire more information because the dimension of DI-QKD's quantum systems is arbitrary. Then the following inequality will hold,
\begin{equation} \label{dimensional:restriction}
\begin{aligned}
H_{\rm min}^{\rm 2\dim}(A_{1}|E) \geq H_{\rm min}^{\rm DI}(A_{1}|E).
\end{aligned}
\end{equation}
The secure key rate of P2 is given by
\begin{equation} \label{finally key rate 2}
\begin{aligned}
R_{2} =&I_{2}(A_{1}:B_{0})-\chi_{2}(A_{1}:E)\\
=&I_{2}(A_{1}:B_{0})-[\chi_{2}^{\rm tag}(A_{1}:E)+\chi_{2}^{\rm untag}(A_{1}:E)]\\
\geq &Q_{\mu\nu}^{Z}-Q_{\mu\nu}^{Z}fH(E_{\mu\nu}^{Z})-(Q_{\mu\nu}^{Z}-Q_{11}^{Z})\\
&-Q_{11}^{Z}\Big[1+{\rm log_{2}}\left(\frac{1}{2}+\frac{1}{2}\sqrt{2-\frac{S_{11}^2}{4}}\right)\Big]\\
= & Q_{11}^{Z}\Big[1-{\rm log_{2}}\left(1+\sqrt{2-\frac{S_{11}^2}{4}}\right)\Big]-Q_{\mu\nu}^ZfH(E_{\mu\nu}^Z).
\end{aligned}
\end{equation}

\section{ESTIMATE $Q_{11}^{Z}$, $e_{11}^{BZ}$ and $S_{11}$} \label{Appsection:Estimate}
\subsection{gain and error}
Now, we evaluate the overall gain and QBER. Alice and Bob prepare phase-randomized  WCPs with intensity $\mu_{i}$ and $\nu_{j}$, respectively.
The overall gain and QBER in $\sigma_z$ basis (Alice chooses setting $A_{1}$ and Bob chooses setting $B_{0}$) can be written as \cite{Ma:2012:Statistical}
\begin{equation} \label{Gain:ERROR}
\begin{aligned}
Q_{\mu_{i}\nu_{j}}^{Z} = Q_{\mu_{i}\nu_{j}}^{CZ}&+Q_{\mu_{i}\nu_{j}}^{EZ}=\sum_{n=0}^\infty\sum_{m=0}^\infty\frac{\mu_{i}^n\nu_{j}^m}{n!m!}e^{-\mu_{i}-\nu_{j}}Y_{nm}^{Z},\\
E_{\mu_{i}\nu_{j}}^{Z}Q_{\mu_{i}\nu_{j}}^{Z}&= e_{d}Q_{\mu_{i}\nu_{j}}^{CZ}+(1-e_{d})Q_{\mu_{i}\nu_{j}}^{EZ}\\
&=\sum_{n=0}^\infty\sum_{m=0}^\infty\frac{\mu_{i}^n\nu_{j}^m}{n!m!}e^{-\mu_{i}-\nu_{j}}e_{nm}^{BZ}Y_{nm}^{Z},
\end{aligned}
\end{equation}
where
\begin{equation} \label{QCZ}
\begin{aligned}
Q_{\mu_{i}\nu_{j}}^{CZ}= &2(1-p_{d})^2e^{-\frac{\omega}{2}}\big[1-(1-p_{d})e^{-\frac{\mu_{i}\eta_{a}}{2}}\big]\\
&\times\big[1-(1-p_{d})e^{-\frac{\nu_{j}\eta_{b}}{2}}\big],\\
Q_{\mu_{i}\nu_{j}}^{EZ}= &2p_{d}(1-p_{d})^2e^{-\frac{\omega}{2}}\big[I_{0}(2x)-(1-p_{d})e^{-\frac{\omega}{2}}\big].\\
\end{aligned}
\end{equation}
In the above equations, $p_{d}$ is the background count rate, $I_{0}(2x)$ is the modified Bessel function of the first kind, $e_{d}$ represents the misalignment-error probability, and $\omega=\mu_{i}\eta_{a}+\nu_{j}\eta_{b}$, $x=\frac{\sqrt{\mu_{i}\nu_{j}\eta_{a}\eta_{b}}}{2}$. $\eta_{a}=\eta_{b}=\eta_{d}\times10^{-\beta L/20}$ is the total efficiency including channel transmittance efficiency $10^{-\beta L/20}$ and detection efficiency $\eta_{d}$. Considering the symmetric scenario, the distance between Alice (Bob) and Charlie is $L/2$.

Now, we focus on the joint quantum state. Alice sends out a positive eigenvalue corresponding to the eigenstate $\ket{H_{A_{1}}}=\ket{H}$ of setting $A_{1}$ and Bob sends out a positive eigenvalue corresponding to the eigenstate $\ket{H_{B_{1}}}=\cos{\frac{\pi}{8}}\ket{H}+\cos{\frac{3\pi}{8}}\ket{V}$ of setting $B_{1}$, i.e.,
\begin{equation} \label{quantum state in}
\begin{aligned}
\ket{H_{A_{1}}}\otimes\ket{H_{B_{1}}}=&\ket{e^{i\phi_{a}}\sqrt{\mu_{i}\eta_{a}}}_{H}\otimes\big(\cos{\frac{\pi}{8}}\ket{e^{i\phi_{b}}\sqrt{\nu_{j}\eta_{b}}}_{H}\\
&+\cos{\frac{3\pi}{8}}\ket{e^{i\phi_{b}}\sqrt{\nu_{j}\eta_{b}}}_{V}\big),
\end{aligned}
\end{equation}
where $\phi_{a}$ and $\phi_{b}$ are the overall randomized phases, while $\ket{H}$ ($\ket{V}$) is a positive (negative) eigenvalue corresponding to the eigenstate of $\sigma_z$ basis. Then the quantum state passing through the beam splitter and four polarization beam splitters is given by
\begin{equation} \label{quantum state out}
\begin{aligned}
&\ket{e^{i\phi_{a}}\sqrt{\frac{\mu_{i}\eta_{a}}{2}}+\cos{\frac{\pi}{8}}e^{i\phi_{b}}\sqrt{\frac{\nu_{j}\eta_{b}}{2}}}_{1H}\ket{\cos{\frac{3\pi}{8}}e^{i\phi_{b}}\sqrt{\frac{\nu_{j}\eta_{b}}{2}}}_{1V}\\
&\quad\otimes\ket{e^{i\phi_{a}}\sqrt{\frac{\mu_{i}\eta_{a}}{2}}-\cos{\frac{\pi}{8}}e^{i\phi_{b}}\sqrt{\frac{\nu_{j}\eta_{b}}{2}}}_{2H}\\
&\quad\otimes\ket{-\cos{\frac{3\pi}{8}}e^{i\phi_{b}}\sqrt{\frac{\nu_{j}\eta_{b}}{2}}}_{2V},
\end{aligned}
\end{equation}
where the four detection modes are $1H$, $1V$, $2H$ and $2V$. Therefore, the detection probabilities for the four detectors are given by
\begin{equation} \label{detector probability}
\begin{aligned}
D_{1H}&=1-(1-p_{d})\exp(-|\frac{e^{i\phi_{a}}\sqrt{\mu_{i}\eta_{a}}+\cos\frac{\pi}{8}e^{i\phi_{b}}\sqrt{\nu_{j}\eta_{b}}}{\sqrt{2}}|^2),\\
D_{1V}&=1-(1-p_{d})\exp(-|\frac{\cos\frac{3\pi}{8}e^{i\phi_{b}}\sqrt{\nu_{j}\eta_{b}}}{\sqrt{2}}|^2),\\
D_{2H}&=1-(1-p_{d})\exp(-|\frac{e^{i\phi_{a}}\sqrt{\mu_{i}\eta_{a}}-\cos\frac{\pi}{8}e^{i\phi_{b}}\sqrt{\nu_{j}\eta_{b}}}{\sqrt{2}}|^2),\\
D_{2V}&=1-(1-p_{d})\exp(-|\frac{-\cos\frac{3\pi}{8}e^{i\phi_{b}}\sqrt{\nu_{j}\eta_{b}}}{\sqrt{2}}|^2).
\end{aligned}
\end{equation}
The gain $Q_{H_{A_{1}}H_{B_{1}}}^{\mu_{i}\nu_{j}\psi^-}$ is defined as the probability that Alice sends out a positive eigenvalue corresponding to the eigenstate $\ket{H_{A_{1}}}=\ket{H}$ of setting $A_{1}$ with the intensity $\mu_{i}$, while Bob sends out a positive eigenvalue corresponding to the eigenstate $\ket{H_{B_{1}}}=\cos{\frac{\pi}{8}}\ket{H}+\cos{\frac{3\pi}{8}}\ket{V}$ of setting $B_{1}$ with the intensity $\nu_{j}$. Meanwhile, Charlie has a successful Bell state $\ket{\psi^-}$ measurement event. Therefore,
\begin{equation} \label{gain:HA1:HB1}
\begin{aligned}
Q_{H_{A_{1}}H_{B_{1}}}^{\mu_{i}\nu_{j}\psi^-}=&\frac{1}{2\pi}\int_0^{2\pi}\frac{1}{4}\big[D_{1H}D_{2V}(1-D_{2H})(1-D_{1V})\\
&+D_{2H}D_{1V}(1-D_{1H})(1-D_{2V})\big]d\phi,
\end{aligned}
\end{equation}
where $Q_{H_{A_{1}}H_{B_{1}}}^{\mu_{i}\nu_{j}\psi^-}$ is averaged over random phases $\phi_{a}$ and $\phi_{b}$, $\phi=\phi_{a}-\phi_{b}$. By substituting Eq.~\eqref{detector probability} into  Eq.~\eqref{gain:HA1:HB1}, we have
\begin{equation} \label{Qain:HA1:HB1}
\begin{aligned}
Q_{H_{A_{1}}H_{B_{1}}}^{\mu_{i}\nu_{j}\psi^-}=&\sum_{n=0}^\infty\sum_{m=0}^\infty\frac{\mu_{i}^n\nu_{j}^m}{n!m!}e^{-\mu_{i}-\nu_{j}}Y_{H_{A_{1}}H_{B_{1}}}^{nm\psi^-}\\
=&\frac{1}{2}(1-p_{d})^2e^{-\frac{\omega}{2}}I_{0}(2x\cos\frac{\pi}{8})+\frac{1}{2}(1-p_{d})^4e^{-\omega}\\
&-\frac{1}{2}(1-p_{d})^3e^{-\frac{2\mu_{i}\eta_{a}+(1+\cos^2\frac{\pi}{8})\nu_{j}\eta_{b}}{2}}\\
&-\frac{1}{2}(1-p_{d})^3e^{-\frac{\mu_{i}\eta_{a}+(1+\cos^2\frac{3\pi}{8})\nu_{j}\eta_{b}}{2}}I_{0}(2x\cos\frac{\pi}{8}).
\end{aligned}
\end{equation}
According to the above procedures, we can also obtain
\begin{equation} \label{Qain:others}
\begin{aligned}
Q_{H_{A_{1}}V_{B_{1}}}^{\mu_{i}\nu_{j}\psi^-}&=\sum_{n=0}^\infty\sum_{m=0}^\infty\frac{\mu_{i}^n\nu_{j}^m}{n!m!}e^{-\mu_{i}-\nu_{j}}Y_{H_{A_{1}}V_{B_{1}}}^{nm\psi^-}\\
&=\frac{1}{2}(1-p_{d})^2e^{-\frac{\omega}{2}}I_{0}(2x\cos\frac{3\pi}{8})+\frac{1}{2}(1-p_{d})^4e^{-\omega}\\
&-\frac{1}{2}(1-p_{d})^3e^{-\frac{2\mu_{i}\eta_{a}+(1+\cos^2\frac{3\pi}{8})\nu_{j}\eta_{b}}{2}}\\
&-\frac{1}{2}(1-p_{d})^3e^{-\frac{\mu_{i}\eta_{a}+(1+\cos^2\frac{\pi}{8})\nu_{j}\eta_{b}}{2}}I_{0}(2x\cos\frac{3\pi}{8}),\\
\end{aligned}
\end{equation}
\begin{equation} \label{Qain:others}
\begin{aligned}
Q_{H_{A_{2}}H_{B_{1}}}^{\mu_{i}\nu_{j}\psi^-}&=\sum_{n=0}^\infty\sum_{m=0}^\infty\frac{\mu_{i}^n\nu_{j}^m}{n!m!}e^{-\mu_{i}-\nu_{j}}Y_{H_{A_{2}}H_{B_{1}}}^{nm\psi^-}\\
&=\frac{1}{2}(1-p_{d})^2e^{-\frac{\omega}{2}}I_{0}(\sqrt{2}x(\cos\frac{\pi}{8}-\cos\frac{3\pi}{8}))\\
&-\frac{1}{2}(1-p_{d})^3e^{-\frac{\frac{3}{2}\mu_{i}\eta_{a}+(1+\cos^2\frac{\pi}{8})\nu_{j}\eta_{b}}{2}}I_{0}(\sqrt{2}x\cos\frac{3\pi}{8})\\
&-\frac{1}{2}(1-p_{d})^3e^{-\frac{\frac{3}{2}\mu_{i}\eta_{a}+(1+\cos^2\frac{3\pi}{8})\nu_{j}\eta_{b}}{2}}I_{0}(\sqrt{2}x\cos\frac{\pi}{8})\\
&+\frac{1}{2}(1-p_{d})^4e^{-\omega},\\
\end{aligned}
\end{equation}
\begin{equation} \label{Qain:others}
\begin{aligned}
Q_{H_{A_{2}}V_{B_{1}}}^{\mu_{i}\nu_{j}\psi^-}&=\sum_{n=0}^\infty\sum_{m=0}^\infty\frac{\mu_{i}^n\nu_{j}^m}{n!m!}e^{-\mu_{i}-\nu_{j}}Y_{H_{A_{2}}V_{B_{1}}}^{nm\psi^-}\\
&=\frac{1}{2}(1-p_{d})^2e^{-\frac{\omega}{2}}I_{0}(\sqrt{2}x(\cos\frac{\pi}{8}+\cos\frac{3\pi}{8}))\\
&-\frac{1}{2}(1-p_{d})^3e^{-\frac{\frac{3}{2}\mu_{i}\eta_{a}+(1+\cos^2\frac{\pi}{8})\nu_{j}\eta_{b}}{2}}I_{0}(\sqrt{2}x\cos\frac{3\pi}{8})\\
&-\frac{1}{2}(1-p_{d})^3e^{-\frac{\frac{3}{2}\mu_{i}\eta_{a}+(1+\cos^2\frac{3\pi}{8})\nu_{j}\eta_{b}}{2}}I_{0}(\sqrt{2}x\cos\frac{\pi}{8})\\
&+\frac{1}{2}(1-p_{d})^4e^{-\omega},
\end{aligned}
\end{equation}
and
\begin{equation} \label{Qain:others1}
\begin{aligned}
Q_{H_{A_{1}}H_{B_{1}}}^{\mu_{i}\nu_{j}\psi^-}&=Q_{V_{A_{1}}V_{B_{1}}}^{\mu_{i}\nu_{j}\psi^-}=Q_{H_{A_{1}}H_{B_{2}}}^{\mu_{i}\nu_{j}\psi^-}=Q_{V_{A_{1}}V_{B_{2}}}^{\mu_{i}\nu_{j}\psi^-},\\
Q_{H_{A_{1}}V_{B_{1}}}^{\mu_{i}\nu_{j}\psi^-}&=Q_{V_{A_{1}}H_{B_{1}}}^{\mu_{i}\nu_{j}\psi^-}=Q_{H_{A_{1}}V_{B_{2}}}^{\mu_{i}\nu_{j}\psi^-}=Q_{V_{A_{1}}H_{B_{2}}}^{\mu_{i}\nu_{j}\psi^-},\\
Q_{H_{A_{2}}H_{B_{1}}}^{\mu_{i}\nu_{j}\psi^-}&=Q_{V_{A_{2}}V_{B_{1}}}^{\mu_{i}\nu_{j}\psi^-}=Q_{H_{A_{2}}V_{B_{2}}}^{\mu_{i}\nu_{j}\psi^-}=Q_{V_{A_{2}}H_{B_{2}}}^{\mu_{i}\nu_{j}\psi^-},\\
Q_{H_{A_{2}}V_{B_{1}}}^{\mu_{i}\nu_{j}\psi^-}&=Q_{V_{A_{2}}H_{B_{1}}}^{\mu_{i}\nu_{j}\psi^-}=Q_{H_{A_{2}}H_{B_{2}}}^{\mu_{i}\nu_{j}\psi^-}=Q_{V_{A_{2}}V_{B_{2}}}^{\mu_{i}\nu_{j}\psi^-}.
\end{aligned}
\end{equation}

\subsection{Asymptotic case}
The gain of single-photon states (untagged portion) in $\sigma_z$ basis, $Q_{11}^{Z}$, is given by
\begin{equation} \label{Q11}
\begin{aligned}
Q_{11}^{Z}=\mu\nu e^{-\mu-\nu}Y_{11}^{Z}.
\end{aligned}
\end{equation}
For the asymptotic case (with infinite number of decoy states and infinite data length), the yield and bit error rate in $\sigma_z$ basis with single-photon states are given by \cite{Ma:2012:Statistical}
\begin{equation} \label{finally key rate}
\begin{aligned}
Y_{11}^Z=&(1-p_{d})^2\big[\frac{\eta_{a}\eta_{b}}{2}+(2\eta_{a}+2\eta_{b}-3\eta_{a}\eta_{b})p_{d}\\
&+4(1-\eta_{a})(1-\eta_{b})p_{d}^2\big],\\
e_{11}^{BZ}Y_{11}^{Z}=&e_{0}Y_{11}^Z-(e_{0}-e_{d})(1-p_{d})^2(1-2p_{d})\frac{\eta_{a}\eta_{b}}{2},
\end{aligned}
\end{equation}
where $e_{0}=\frac{1}{2}$. The Bell value $S_{11}^{\psi^-}$ of single-photon states is given by
\begin{equation} \label{CHSH}
\begin{aligned}
S_{11}^{\psi^-}&=\langle A_{2}B_{2}\rangle_{11}^{\psi^-}-\langle A_{2}B_{1}\rangle_{11}^{\psi^-}-\langle A_{1}B_{2}\rangle_{11}^{\psi^-}-\langle A_{1}B_{1}\rangle_{11}^{\psi^-},
\end{aligned}
\end{equation}
where
\begin{equation} \label{EAB}
\begin{aligned}
\langle A_{k}B_{l}\rangle_{11}^{\psi^-}=&(1-2e_{d})\\
&\times\frac{Y_{H_{A_{k}}H_{B_{l}}}^{11\psi^-}+Y_{V_{A_{k}}V_{B_{l}}}^{11\psi^-}-Y_{H_{A_{k}}V_{B_{l}}}^{11\psi^-}-Y_{V_{A_{k}}H_{B_{l}}}^{11\psi^-}}{Y_{H_{A_{k}}H_{B_{l}}}^{11\psi^-}+Y_{V_{A_{k}}V_{B_{l}}}^{11\psi^-}+Y_{H_{A_{k}}V_{B_{l}}}^{11\psi^-}+Y_{V_{A_{k}}H_{B_{l}}}^{11\psi^-}},
\end{aligned}
\end{equation}
$k,l\in\{1,2\}$. Thereinto,
\begin{equation} \label{yield}
\begin{aligned}
Y_{H_{A_{1}}H_{B_{1}}}^{11\psi^-}=&\cos^2\frac{\pi}{8}\frac{p_{d}}{4}(1-p_{d})^2[1-(1-2p_{d})(1-\eta_{a})\\
&\times(1-\eta_{b})]+\frac{p_{d}}{8}\cos^2\frac{3\pi}{8}(1-p_{d})^2\\
&\times[(2-\eta_{a}-\eta_{b})+2(1-p_{d})(1-\eta_{a})(1-\eta_{b})]\\
&+\frac{(1-p_{d})^2}{8}\cos^2\frac{3\pi}{8}[p_{d}(\eta_{a}+\eta_{b})\\
&+(1-2p_{d})\eta_{a}\eta_{b}+2p_{d}^2(1-\eta_{a})(1-\eta_{b})],\\
\end{aligned}
\end{equation}
\begin{equation} \label{yield}
\begin{aligned}
Y_{H_{A_{1}}V_{B_{1}}}^{11\psi^-}=&\cos^2\frac{3\pi}{8}\frac{p_{d}}{4}(1-p_{d})^2[1-(1-2p_{d})(1-\eta_{a})\\
&\times(1-\eta_{b})]+\frac{p_{d}}{8}\cos^2\frac{\pi}{8}(1-p_{d})^2\\
&\times[(2-\eta_{a}-\eta_{b})+2(1-p_{d})(1-\eta_{a})(1-\eta_{b})]\\
&+\frac{(1-p_{d})^2}{8}\cos^2\frac{\pi}{8}[p_{d}(\eta_{a}+\eta_{b})\\
&+(1-2p_{d})\eta_{a}\eta_{b}+2p_{d}^2(1-\eta_{a})(1-\eta_{b})],\\
\end{aligned}
\end{equation}
\begin{equation} \label{yield}
\begin{aligned}
Y_{H_{A_{2}}H_{B_{1}}}^{11\psi^-}=&\frac{p_{d}}{8}(1-p_{d})^2[1-(1-2p_{d})(1-\eta_{a})(1-\eta_{b})]\\
&+\frac{p_{d}}{8}(\cos\frac{\pi}{8}+\cos\frac{3\pi}{8})^2(1-p_{d})^2[(2-\eta_{a}-\eta_{b})\\
&+2(1-p_{d})(1-\eta_{a})(1-\eta_{b})]\\
&+\frac{(1-p_{d})^2}{16}(\cos\frac{\pi}{8}-\cos\frac{3\pi}{8})^2[p_{d}(\eta_{a}+\eta_{b})\\
&+(1-2p_{d})\eta_{a}\eta_{b}+2p_{d}^2(1-\eta_{a})(1-\eta_{b})],\\
\end{aligned}
\end{equation}
\begin{equation} \label{yield}
\begin{aligned}
Y_{H_{A_{2}}V_{B_{1}}}^{11\psi^-}&=\frac{p_{d}}{8}(1-p_{d})^2[1-(1-2p_{d})(1-\eta_{a})(1-\eta_{b})]\\
&+\frac{p_{d}}{8}(\cos\frac{\pi}{8}-\cos\frac{3\pi}{8})^2(1-p_{d})^2[(2-\eta_{a}-\eta_{b})\\
&+2(1-p_{d})(1-\eta_{a})(1-\eta_{b})]\\
&+\frac{(1-p_{d})^2}{16}(\cos\frac{\pi}{8}+\cos\frac{3\pi}{8})^2[p_{d}(\eta_{a}+\eta_{b})\\
&+(1-2p_{d})\eta_{a}\eta_{b}+2p_{d}^2(1-\eta_{a})(1-\eta_{b})],
\end{aligned}
\end{equation}
and
\begin{equation} \label{Asymptotic Yield}
\begin{aligned}
Y_{H_{A_{1}}H_{B_{1}}}^{11\psi^-}=Y_{V_{A_{1}}V_{B_{1}}}^{11\psi^-}=Y_{H_{A_{1}}H_{B_{2}}}^{11\psi^-}=Y_{V_{A_{1}}V_{B_{2}}}^{11\psi^-},\\
Y_{H_{A_{1}}V_{B_{1}}}^{11\psi^-}=Y_{V_{A_{1}}H_{B_{1}}}^{11\psi^-}=Y_{H_{A_{1}}V_{B_{2}}}^{11\psi^-}=Y_{V_{A_{1}}H_{B_{2}}}^{11\psi^-},\\
Y_{H_{A_{2}}H_{B_{1}}}^{11\psi^-}=Y_{V_{A_{2}}V_{B_{1}}}^{11\psi^-}=Y_{H_{A_{2}}V_{B_{2}}}^{11\psi^-}=Y_{V_{A_{2}}H_{B_{2}}}^{11\psi^-},\\
Y_{H_{A_{2}}V_{B_{1}}}^{11\psi^-}=Y_{V_{A_{2}}H_{B_{1}}}^{11\psi^-}=Y_{H_{A_{2}}H_{B_{2}}}^{11\psi^-}=Y_{V_{A_{2}}V_{B_{2}}}^{11\psi^-}.
\end{aligned}
\end{equation}

\subsection{Finite decoy-state case}
In practical demonstrations, the length of the raw key is finite, which will induce statistical fluctuations for the parameter estimation. Here, we consider the effect of finite length raw key based on standard error analysis method \cite{Ma:2005:Practical,Ma:2012:Statistical}. The estimations of $Y_{11}^{Z}$, $e_{11}^{BZ}$ and $S_{11}^{\psi^-}$ are constrained optimization problems, which are linear and can be efficiently solved by linear programming \cite{Ma:2012:Statistical,Xu:2014:Protocol}.

Now, we consider an analytical estimation method with two decoy states \cite{Xu:2013:practical}, $\mu_{2}=\nu_{2}>\mu_{1}=\nu_{1}>\mu_{0}=\nu_{0}=0$. The lower bound of $Y_{11}^{ZL}$ , the upper bound of $Y_{11}^{ZU}$ and the lower bound of $e_{11}^{BZL}$ are given by
\begin{equation} \label{yield}
\begin{aligned}
Y_{11}^{ZL}\geq&\frac{1}{\mu_{2}^2\mu_{1}^2(\mu_{2}-\mu_{1})}\Big[\mu_{2}^3\big(e^{2\mu_{1}}Q_{\mu_{1}\mu_{1}}^{Z}+Q_{00}^{Z}\\
&-e^{\mu_{1}}Q_{\mu_{1}0}^{Z}-e^{\mu_{1}}Q_{0\mu_{1}}^{Z}\big)-\mu_{1}^3\big(e^{2\mu_{2}}Q_{\mu_{2}\mu_{2}}^{Z}\\
&+Q_{00}^{Z}-e^{\mu_{2}}Q_{\mu_{2}0}^{Z}-e^{\mu_{2}}Q_{0\mu_{2}}^{Z}\big)\Big],
\end{aligned}
\end{equation}
\begin{equation} \label{yield}
\begin{aligned}
Y_{11}^{ZU}\leq&\frac{1}{\mu_{1}^{2}}\big[e^{2\mu_{1}}Q_{\mu_{1}\mu_{1}}^{Z}+Q_{00}^{Z}-e^{\mu_{1}}Q_{\mu_{1}0}^{Z}-e^{\mu_{1}}Q_{0\mu_{1}}^{Z}\big],
\end{aligned}
\end{equation}
\begin{equation} \label{error}
\begin{aligned}
e_{11}^{BZL}\geq&\frac{1}{\mu_{2}^2\mu_{1}^2(\mu_{2}-\mu_{1})Y_{11}^{ZU}}\Bigg\{\mu_{2}^3\Big[e^{2\mu_{1}}E_{\mu_{1}\mu_{1}}^{Z}Q_{\mu_{1}\mu_{1}}^{Z}\\
&+E_{00}^{Z}Q_{00}^{Z}-e^{\mu_{1}}E_{\mu_{1}0}^{Z}Q_{\mu_{1}0}^{Z}-e^{\mu_{1}}E_{0\mu_{1}}^{Z}Q_{0\mu_{1}}^{Z}\Big]\\
&-\mu_{1}^3\Big[e^{2\mu_{2}}E_{\mu_{2}\mu_{2}}^{Z}Q_{\mu_{2}\mu_{2}}^{Z}+E_{00}^{Z}Q_{00}^{Z}\\
&-e^{\mu_{2}}E_{\mu_{2}0}^{Z}Q_{\mu_{2}0}^{Z}-e^{\mu_{2}}E_{0\mu_{2}}^{Z}Q_{0\mu_{2}}^{Z}\Big]\Bigg\}.
\end{aligned}
\end{equation}
Combining Eq.~\eqref{Qain:others1} and Eq.~\eqref{Asymptotic Yield}, we can use the following equations to estimate the lower bound of $S_{11}^{\psi^-}$,
\begin{equation} \label{S11L}
\begin{aligned}
S_{11}^{\psi^-L}\geq&2(1-2e_{d})\\
&\times\left(\frac{Y_{H_{a1}V_{b1}}^{11\psi^-L}-Y_{H_{A_{1}}H_{B_{1}}}^{11\psi^-U}}{Y_{H_{A_{1}}H_{B_{1}}}^{11\psi^-U}+Y_{H_{A_{1}}V_{B_{1}}}^{11\psi^-U}}+\frac{Y_{H_{A_{2}}V_{B_{1}}}^{11\psi^-L}-Y_{H_{A_{2}}H_{B_{1}}}^{11\psi^-U}}{Y_{H_{A_{2}}H_{B_{1}}}^{11\psi^-U}+Y_{H_{A_{2}}V_{B_{1}}}^{11\psi^-U}}\right),
\end{aligned}
\end{equation}
where
\begin{equation} \label{estimate yield}
\begin{aligned}
Y_{H_{A_{k}}V_{B_{l}}}^{11\psi^-L}\geq&\frac{1}{\mu_{2}^2\mu_{1}^2(\mu_{2}-\mu_{1})}\Big[\mu_{2}^3\big(e^{2\mu_{1}}Q_{H_{A_{k}}V_{B_{l}}}^{\mu_{1}\mu_{1}\psi^-}+Q_{H_{A_{k}}V_{B_{l}}}^{00\psi^-}\\
&-e^{\mu_{1}}Q_{H_{A_{k}}V_{B_{l}}}^{\mu_{1}0\psi^-}-e^{\mu_{1}}Q_{H_{A_{k}}V_{B_{l}}}^{0\mu_{1}\psi^-}\big)\\
&-\mu_{1}^3\big(e^{2\mu_{2}}Q_{H_{A_{k}}V_{B_{l}}}^{\mu_{2}\mu_{2}\psi^-}+Q_{H_{A_{k}}V_{B_{l}}}^{00\psi^-}\\
&-e^{\mu_{2}}Q_{H_{A_{k}}V_{B_{l}}}^{\mu_{2}0\psi^-}-e^{\mu_{2}}Q_{H_{A_{k}}V_{B_{l}}}^{0\mu_{2}\psi^-}\big)\Big],\\
\end{aligned}
\end{equation}
\begin{equation} \label{estimate yield}
\begin{aligned}
Y_{H_{A_{k}}H_{B_{l}}}^{11\psi^-U}\leq&\frac{1}{\mu_{1}^2}\Big[e^{2\mu_{1}}Q_{H_{A_{k}}H_{B_{l}}}^{\mu_{1}\mu_{1}\psi^-}+Q_{H_{A_{k}}H_{B_{l}}}^{00\psi^-}\\
&-e^{\mu_{1}}Q_{H_{A_{k}}H_{B_{l}}}^{\mu_{1}0\psi^-}-e^{\mu_{1}}Q_{H_{A_{k}}H_{B_{l}}}^{0\mu_{1}\psi^-}\Big],\\
\end{aligned}
\end{equation}
\begin{equation} \label{estimate yield}
\begin{aligned}
Y_{H_{A_{k}}V_{B_{l}}}^{11\psi^-U}\leq&\frac{1}{\mu_{1}^2}\Big[e^{2\mu_{1}}Q_{H_{A_{k}}V_{B_{l}}}^{\mu_{1}\mu_{1}\psi^-}+Q_{H_{A_{k}}V_{B_{l}}}^{00\psi^-}\\
&-e^{\mu_{1}}Q_{H_{A_{k}}V_{B_{l}}}^{\mu_{1}0\psi^-}-e^{\mu_{1}}Q_{H_{A_{k}}V_{B_{l}}}^{0\mu_{1}\psi^-}\Big],
\end{aligned}
\end{equation}
and $k,l\in\{1,2\}$.

\end{appendix}
%%%%%%%%%%%%%%%%%%%%%%%%%%%%%%%%%%%%%%%%
% choose a style
%\bibliographystyle{ieeetr}
%\bibliographystyle{unsrt}
%%%%%%%%%%%%%%%%%%%%%%%%%%%%%%%%%%%%%%%%

%%%%%%%%%%%%%%%%%%%%%%%%%%%%%%%%%%%%%%%%
% choose a .bib file
\bibliographystyle{apsrev4-1}
%\bibliography{Biblitex}
%%%%%%%%%%%%%%%%%%%%%%%%%%%%%%%%%%%%%%%%

%\nocite{*}S

\end{document}